# Characterization of solution processed TiO$_x$ buffer layers in inverted organic photovoltaics by XPS and DFT studies

Ivan S. Zhidkov[1,2], Ernst Z. Kurmaev[1,*], Michel A. Korotin[1], Andrey I. Kukharenko[1,2], Achilleas Savva[3], Stelios A. Choulis[3], Danila M. Korotin[1] and Seif O. Cholakh[2]

[1] Institute of Metal Physics, Russian Academy of Sciences – Ural Division, S. Kovalevskoi Str. 18, 620990 Yekaterinburg, Russia

[2] Institute of Physics and Technology, Ural Federal University, Mira Str. 19, 620002 Yekaterinburg, Russia

[3] Molecular Electronics and Photonics Research Unit, Department of Mechanical Engineering and Materials Science and Engineering, Cyprus University of Technology, Kitiou Kiprianou str. 45, 3603 Limassol, Cyprus.

* Corresponding author

E-mail: kurmaev@ifmlrs.uran.ru  (Ernst Z. Kurmaev)

**Abstract**

We present the results of XPS measurements (core levels and valence bands) of solution processed $TiO_x$ thin films prepared from titanium butoxide ($C_{16}H_{36}O_4Ti$) diluted in isopropanol which is a common sol-gel route for fabricating $TiO_x$ electron selective contacts for ITO/TiOx inverted organic photovoltaics bottom electrodes. XPS Ti *2p* and valence band spectra show the presence of additional features which are absent in spectra of titanium butoxide deposited on Si and are attributed to appearance of $Ti^{3+}$ valence states in $ITO/TiO_x$. This conclusion is confirmed by density functional theory electronic structure calculations of stoichiometric $TiO_2$ and oxygen deficient $TiO_{2-1/8}$. XPS C *1s* measurements show the formation of C-O and O-C=O bonds which evidence the presence of residual carbon which can draw oxygen from the film network and induce the formation of fraction of $Ti^{3+}$ states in $TiO_x$ films.

Keywords: $TiO_x$ buffer layers, X-ray photoelectron spectroscopy, density functional theory, sol-gel $TiO_x$, organic optoelectronics, inverted organic photovoltaics

PACS: 88.40.jr, 71.15.Mb, 79.60.Jv

# 1. Introduction

Organic semiconductors are of increasing interest as new materials for opto-electronic devices. This class of semiconductors is processed from solution, offering potential for low fabrication costs and flexibility. Organic light emitting diodes (OLEDs), photodiodes and solar cells are the opto-electronic applications under intense study. Solution processed $TiO_x$, has found an extensive utility as electronic material in organic and hybrid optoelectronic applications [1, 2]. In particular the development of alternative methods for energy production in the past decade has received an increased attention. One of the most attractive in terms of sustainability and development prospects appear a direct conversion of solar energy into electrical current. To date several areas of solar energy materials development are formed, one of which is the creation of solar cells based on hybrid and organic semiconductors. TiOx has been used as electron acceptor in hybrid solar cells [3-5].

More recently solution processed $TiO_x$ has been introduced as an efficient electron selective contact in inverted organic photovoltaics (OPVs) [6]. Inverted structured solution based photovoltaics (PVs), which are based on bottom cathode/top anode electrodes, received scientific interest during the last years due to interesting interfaces created within this structure [7]. Efficient inverted OPVs require highly transparent and conductive materials, which are inserted between the transparent conductive bottom electrode and photoative layer to provide appropriate electron carrier selectivity [8].

One of the biggest tasks in the design of economically viable PVs, is the choice of reliable, environmental robust and versatile processable electron-selective layers. Based on the latter, metal oxides due to their excellent optoelectronic properties, low fabrication cost and chemical/moisture resistance, have been extensively used in cathode electrode configurations, for solution processed OPVs [9]. The most common materials used as electron selective layers, between the bottom transparent contact and the active layer of inverted OPVs are n-type transition metal oxides, like $TiO_x$ [10] and ZnO [11], mainly produced by modified sol-gel processes [12]. The typical schematic diagram of the inverted organic solar cell using TiO$_x$ as electron selective layer between indium tin oxide (ITO) electrode and P3HT/PCBM active layer is shown in Fig. 1 [7].

Although solution processed TiO$_x$ has been used extensively used as electron selective contact in Inverted OPVs there are still open questions related to functionality of the TiO$_x$ buffer layers relevant to device performance which are connected with control of chemical composition of TiO$_x$. In the present paper XPS measurements of were employed to examine the oxidation states of the Ti species in ITO/TiO$_x$ bottom electrode of inverted OPVs . XPS spectra of reference compounds TiO$_2$ and InSnO are also measured. Unlike to the previous XPS studies of ITO/TiO$_x$ in inverted polymer solar cells [13-16] we have measured not only spectra of core levels but also the valence bands which are compared with specially performed density functional theory calculations of stoichiometric and oxygen deficient TiO$_2$. For the XPS measurements, sol-gel prepared TiO$_x$ films have been compared with TiO$_x$ electron selective contacts (produced identically) extracted from high performance inverted OPV devices.

## 2. Sample preparation and experimental details

### 2.1. Sample preparation

Inverted OPVs using TiOx electron selective layer have been fabricated as described in details in previous studies [7, 11]. We have studied XPS spectra of a sol gel prepared TiOx thin films, starting from titanium butoxide precursor coated on glass/ITO substrates (ITO/TiOx) and annealed at 140°C in air and ITO/TiOx films produced identically extracted from a high performing inverted OPV device. The starting material for the preparation of TiO$_x$ buffer layer was Titanium Butoxide (C$_{16}$H$_{36}$O$_4$Ti) diluted in isopropanol. For annealed TiO$_x$ films the precursor film has been annealed at 140°C for 25 minutes in air. For extracted TiO$_x$ films the layers have been extracted from an over 3% working organic photovoltaic (OPV) device (Fig. 1- ITO/TiO$_x$/P3HT:PCBM/PEDOT:PSS/Ag). By using a scotch tape all the layers above TiO$_x$ have been removed and used for the reported XPS measurements.

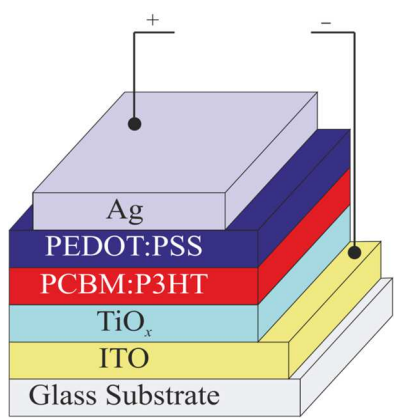

Fig. 1. Schematic diagram of the inverted OPVs used in this study, using TiO$_x$ as electron selective contact [7].

## 2.2. Experimental techniques

XPS core-level and valence band (VB) measurements were made using a PHI 5000 Versa Probe XPS spectrometer (ULVAC Physical Electronics, USA) based on a classic X-ray optic scheme with a hemispherical quartz monochromator and an energy analyzer working in the range of binding energies from 0 to 1500 eV. This apparatus uses electrostatic focusing and magnetic screening to achieve an energy resolution of $\Delta E \leq 0.5$ eV for Al K$\alpha$ radiation (1486.6 eV). Pumping analytical chamber was carried out using an ion pump providing pressure better than $10^{-7}$ Pa. Dual channel neutralization used to local surface charge compensate generated during the measurement. The XPS spectra were recorded using Al K$\alpha$ X-ray emission; the spot size was 200 µm, and the X-ray power load delivered to the sample was less than 50 W. Typical signal-to-noise ratios were greater than 10000:3. Surface cleaning was carried out by Ar$^+$ ions. Simultaneous neutralization of the electron beam was carried out to reduce the influence of the charging effect in the surface layers on XPS spectra. Finally, the spectra were processed using ULVAC-PHI MultiPak Software 9.3 and the residual background was removed using the Shirley method [17]. The XPS spectra were calibrated using reference energy of 285.0 eV for the carbon *1s* core level [18].

## 2.3. DFT calculations

The electronic structure of stoichiometric TiO$_2$ with the rutile structure was calculated using linearized muffin tin orbitals in the atomic sphere and tight bonding approximations (the Stuttgart TB LMTO ASA codes (version 47) [19]) with the experimentally determined lattice constants and atomic positions [20]. We used the following default values for the radii of the MT spheres: $R_{Ti}$ = 2.42 au, $R_O$ = 1.85 au, as well as $R_{ES}$ =1.89 au and 1.71 au (two types of empty spheres, ES). The type of exchange-correlation potential used corresponds to the local density approximation (LDA). As the basis functions, *4s*, *4p*, *3d* states of Ti, *2s*, *2p*, *3d* states of O, and *1s*, *2p*, *3d* states of ES were used. The electronic structure of nonstoichiometric TiO$_{2-\delta}$ ($\delta$=1/8) was calculated within coherent potential approximation CPA [21]. In the CPA calculation scheme, it is assumed that the vacancies are uniformly distributed over all oxygen sites. This means that the lattice symmetry remains unchanged. Within a narrow range of vacancy concentrations, it was also assumed that the lattice parameters do not change in the first approximation.

### 3. Results and discussion

Inverted OPVs have been fabricated using TiO$_x$ as electron selective contact. Fig. 2 demonstrates the current density vs voltage characteristics under dark and under 100mW/cm$^2$ illumination.

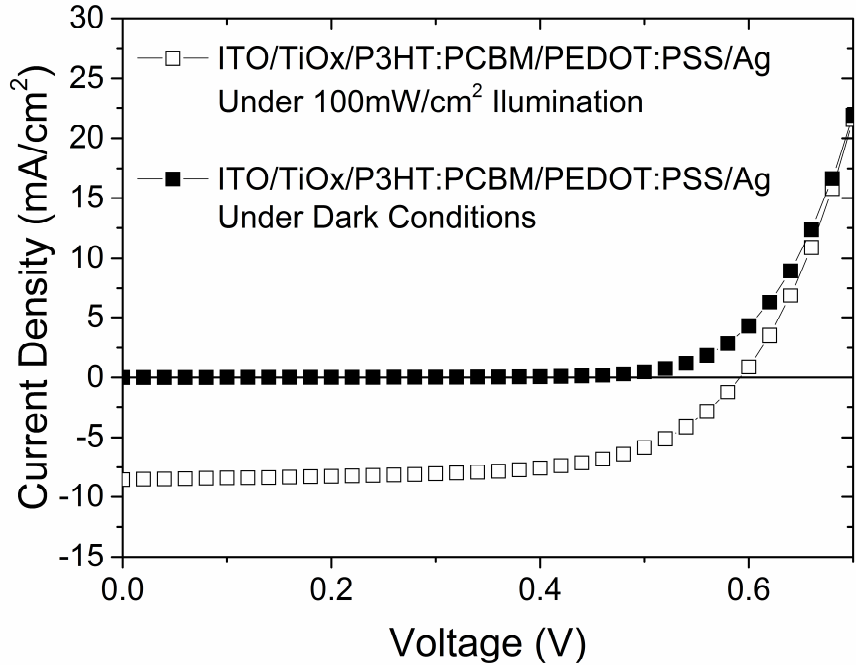

Fig. 2: Current density vs voltage characteristics, under illumination (open squares) and under dark conditions (closed squares), of inverted OPVs using $TiO_x$ as electron selective contact.

Inverted OPVs using $TiO_x$ electron selective layers excibit ideal diode behavior under dark and illuminated conditions. The inverted OPVs under study exhibiting $V_{oc}$=0.591V, $J_{sc}$=8.54mA/cm$^2$, FF=62% and PCE=3.14%. The above observations indicating ideal electron selectivity by $TiO_x$ produced by the proposed sol-gel method. Though, the presented J/V of inverted OPVs have been subjected 10 minutes under UV-light irradiation. Without this light soaking step, inverted OPVs using $TiO_x$ as electron selective layer does not operate efficiently exhibiting a double diode behavior as described by Steim et al [22]. Similar results have been described elsewhere [23].

XPS survey spectra of ITO/$TiO_x$ samples are presented in Fig. 3. One can see that survey spectra consist of signals only from $TiO_2$ protective layer and indium tin oxide without any in-controlled impurities.

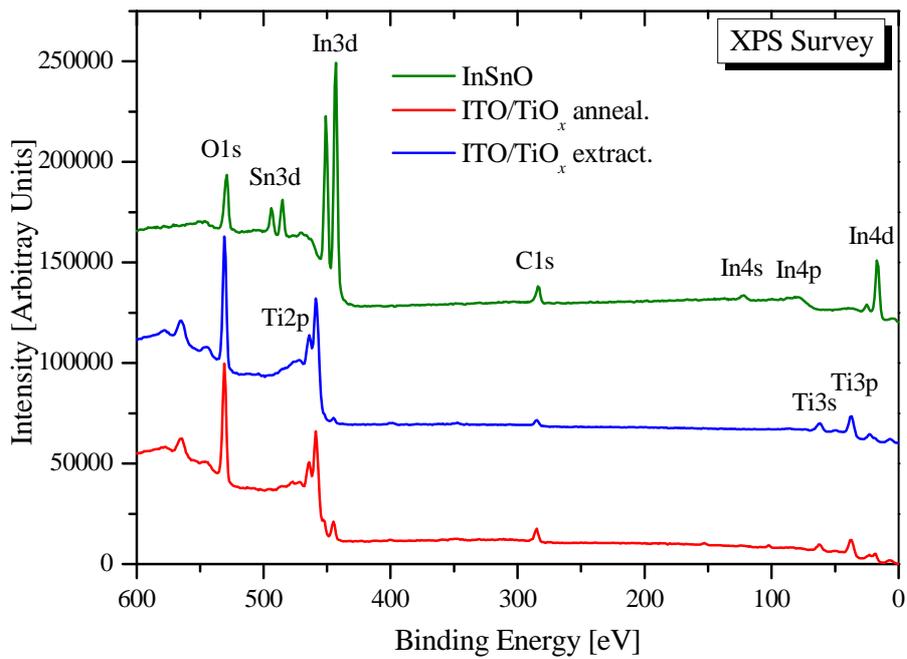

Fig. 3. XPS survey spectra of ITO/TiO$_x$ samples.

The shape and energy position of In$3d$-core levels are found in full accordance with those of InSnO (Fig. 4). As seen from lower panel of Fig. 4 the binding energies and fine structure of In4d and Sn$4d$ subbands are very close to those of InSnO.

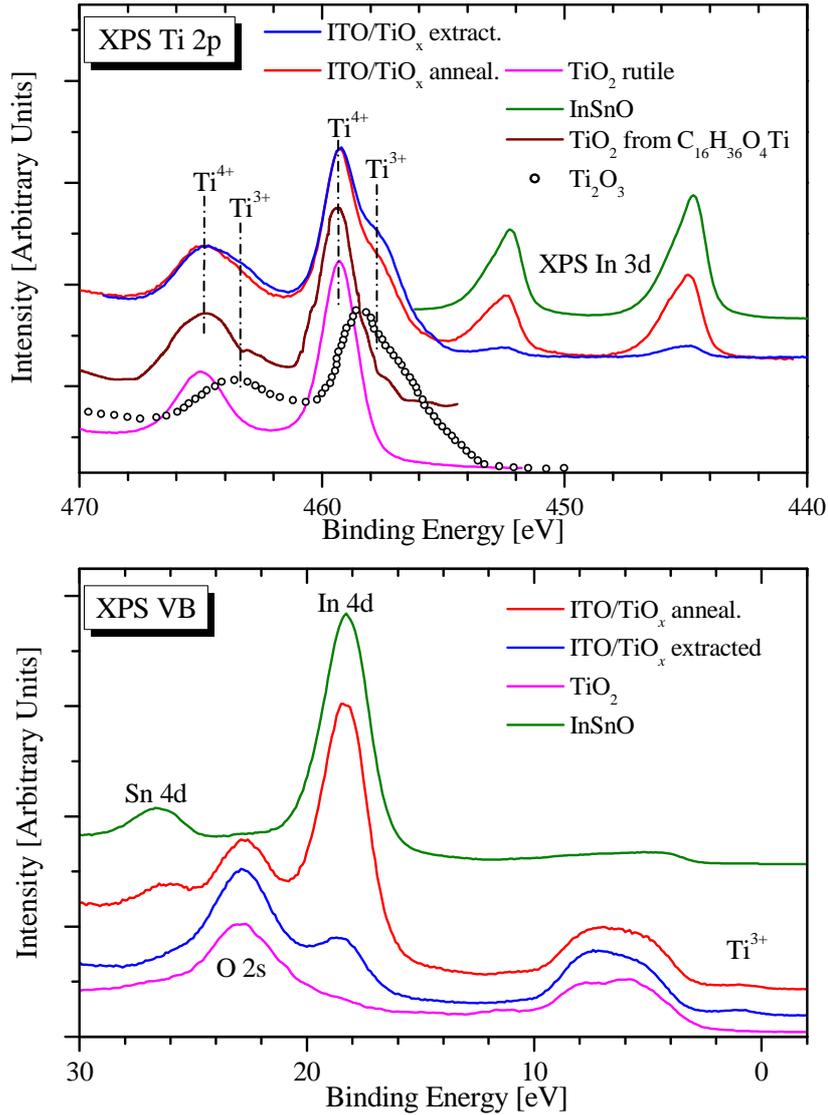

Fig. 4. Upper panel: XPS Ti 2p and In 3d spectra of ITO/TiO$_x$. The spectra of Ti$_2$O$_3$ and TiO$_2$ and Ti butoxide are taken from Refs. [24-25]. Lower panel: XPS VB spectra of ITO/TiO$_x$.

The high-energy resolved XPS Ti*2p* spectra (upper panel of Fig. 4) show the presence of two set of Ti *2p$_{3/2}$* (at 459.3 and 457.8 eV) and *2p$_{1/2}$* (at 464.9 and 463.2 eV) lines which can be attributed to contribution of Ti$^{4+}$ and Ti$^{3+}$ valence states [25-26]. Line corresponding to Ti$^{3+}$ is almost identical in their position and shape to the previously published results for Ti$_2$O$_3$ [25]. These conclusions are confirmed by measurements of XPS valence band spectra which are given in lower panel of Fig. 4 and show that in XPS VB of TiO$_x$ near the Fermi level (0-2 eV) the weak additional feature is revealed which is found in UPS spectra of oxygen-deficient $TiO_2$ [27] and $Ti_2O_3$ [28] and is attributed to appearance of $Ti^{3+}$ states. $Ti^{3+}$ band has a somewhat higher intensity in the ITO/$TiO_x$ extracted from the device than in annealed samples. We also note that the $Ti^{4+}$ band of our samples is closer in structure to a similar line of pure rutile, rather than to previously published results of $TiO_2$ obtained from the Titanium Butoxide [24].

Note the good agreement between the experimental results not only with the literature data, but also with the theoretical calculation. In Fig. 5 XPS valence band spectra of ITO/$TiO_x$ and $TiO_2$ (upper panel) are compared with electronic structure calculations (our data). As seen, the appearance of weak feature at 0-2 eV in experimental spectra can be explained by creation of oxygen vacancies in $TiO_2$ and formation of additional peak in calculated total density of electronic states which is related to $Ti^{3+}$-states. Thus, a combination of experimental and theoretical data clearly shows that solution processed $TiO_x$ electron selective contacts used within inverted OPVs devices revealed the presence of titanium in two oxidation states $Ti^{3+}$ and $Ti^{4+}$.

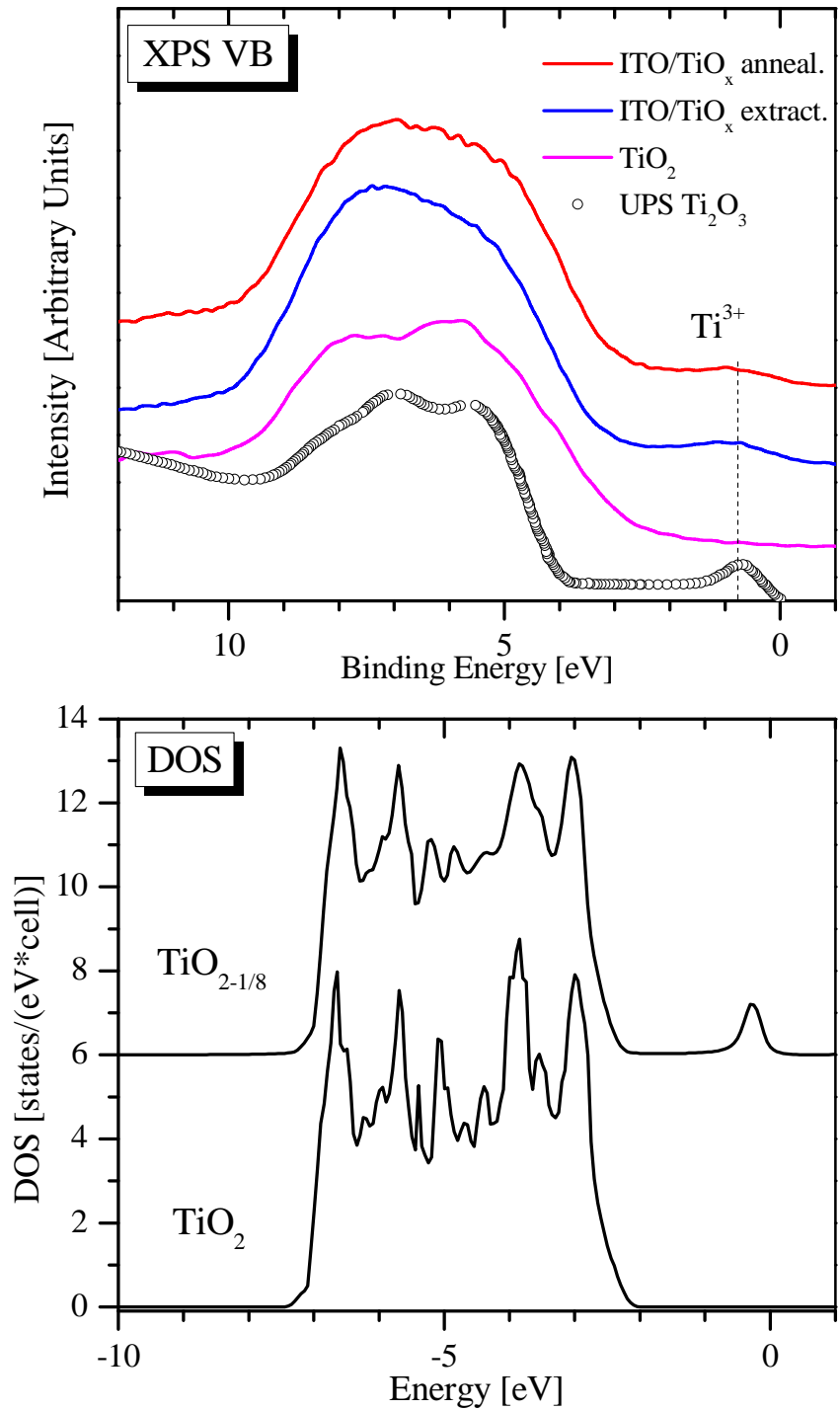

Fig. 5. Upper panel: XPS VB spectra of ITO/TiO$_x$. Lower panel: calculated total density of states of TiO$_2$ and TiO$_{2-1/8}$. UPS spectra are taken from Ref. [28]

On Fig. 5 XPS C $1s$ and O $1s$ spectra are presented. Several bands corresponding to different types of interaction between oxygen and carbon can be distinguished. It is attributed for various units such C–C, C–OH, C–O, C=O, O–C=O [29].

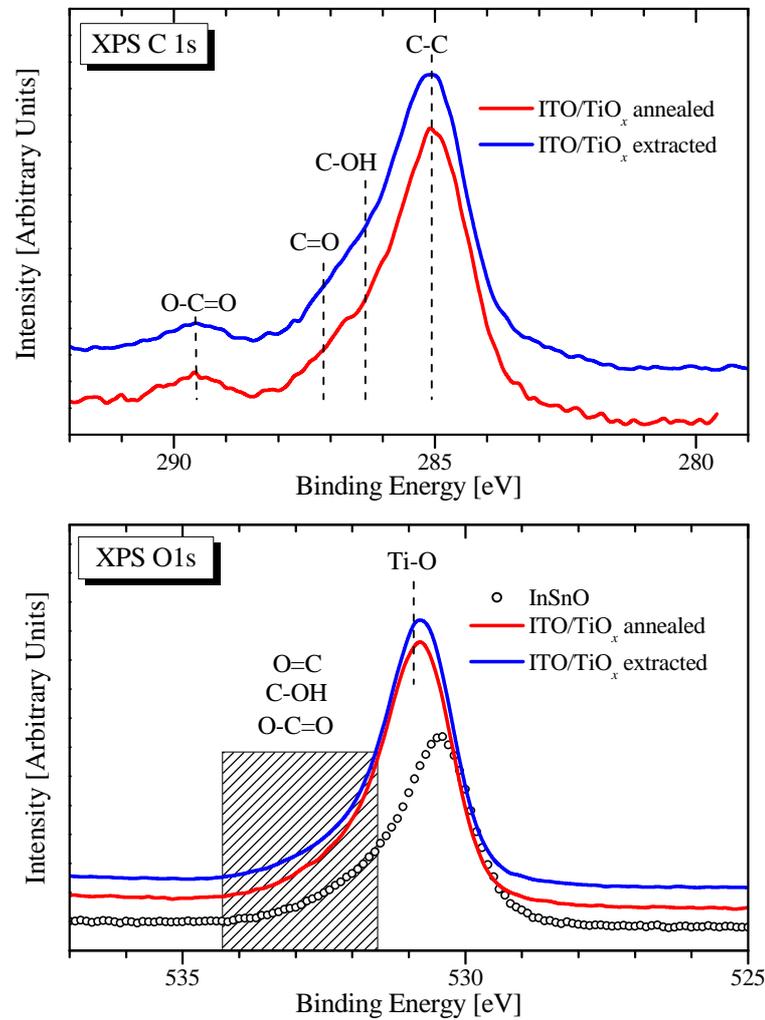

Fig. 6. XPS C1s (upper panel) and O1s (lower panel) spectra of ITO/TiO$_x$.

Going back to discussion of XPS Ti$2p$ spectra (Fig. 4) we need to point out that in spectrum of titanium butoxide deposited on Si no additional low energy features related to Ti$^{3+}$-states are found [24]. What is the reason of their appearance in ITO/TiO$_x$? Generally, the titanium reduction in TiO$_x$ can be induced by the presence of residual carbon which can draw the oxygen from the film network and induce the formation of fraction Ti$^{3+}$ states in TiO$_x$ films [15-16]. It is confirmed by the XPS measurements of C *1s* and O *1s* spectra which indeed show the formation of C–O and O–C=O bonding in ITO/TiO$_x$ films (Fig. 6). This suggestion is also confirmed by our estimation of O/Ti ratio which is found to be rather higher in annealed TiO$_x$ layer than in film extracted from the device (1.8 versus 1.6). This also is consistent with an increase in relative contribution of the Ti$^{3+}$ states in XPS Ti *2p* spectra (Fig. 4, upper panel). Thus in the process of device exploitation the interaction between oxygen and carbon atoms in TiO$_x$ barrier layer is increased. According to previous reports [30-31] Ti$^{3+}$-states in oxygen deficient TiO$_x$ form a donor level just below the conduction band. The Ti$^{3+}$-states can trap the photogenerated electrons and then transfer them to O$_2$ absorbed on the surface of of TiO$_x$ [32]. It has been shown that the existence of Ti$^{3+}$-states in TiO$_x$ electon donors can reduce the rates of electron and hole recombination and enhance the hybrid solar cell performance [3]. As is shown in Ref. 32-33, the modification of surface of TiO$_2$, formation of oxygen vacancies, other defects and Ti$^{3+}$ centers can enhance the TiO$_x$ photocatalytic activity.

## 4. Conclusion

To conclude, we have studied Ti oxidation states in ITO/TiO$_x$ bottom electrode as used within the inverted OPVs device structures. The TiO$_x$ electron selective contact prepared from titanium butoxide (Ti$^{4+}$) diluted in isopropanol. The appearance of Ti$^{3+}$ states in ITO/TiO$_x$ bottom inverted OPV electrodes is proved by the measurements of XPS Ti *2p* and XPS VB spectra and DFT calculations of stoichiometric and oxygen deficient TiO$_2$. The measurements of XPS C *1s* and O *1s* spectra have shown the formation of C–O and O–C=O bonds which can draw of oxygen from the solution processed TiOx buffer layer . By this mechanism the oxygen deficiency in TiO$_x$ electron selective contacts is created and as a result Ti$^{3+}$ surface states form a donor level below the conduction band. The Ti$^{3+}$-states can trap the photogenerated electrons and then transfer them to O$_2$ absorbed on the surface of TiO$_x$, this mechanism can enhance the photocatalytic activity of TiO$_x$ solution processed electron selective contacts used within inverted OPV devices.

**Acknowledgments**

The XPS measurements and DFT calculations are supported by the Russian Scientific Foundation (Project 14-22-00004). Inverted OPVs fabrication and samples preparation was co-funded by the European Regional Development Fund and the Republic of Cyprus through the Research Promotion Foundation (Strategic Infrastructure Project ΝΕΑ ΥΠΟΔΟΜΗ/ΣΤΡΑΤΗ/0308/06).